\documentclass[12pt,draft]{article}

\title{Relational Analysis of Dirac Equation in Momentum Representation}
\author{Anton V. Solov'yov\thanks{Department of Theoretical Physics, Faculty of Physics, M. V. Lomonosov Moscow State University, Leninskie Gory, Moscow, 119991, Russian Federation. E-mail: a.v.solovyov@gmail.com}}
\date{}
\begin{document}
\maketitle
\begin{abstract}
In terms of the relational approach to space-time geometry and physical interactions, we show that the Dirac equation for a free fermion in the momentum representation can be obtained starting from a \textit{binary system of complex relations} (BSCR) between elements of two abstract sets. With the derivation performed we show that the 4-dimensional pseudo-Euclidean momentum space is not needed \textit{a priori} but naturally emerges from considerations of rather general character (2-spinor algebra). A bispinor wave function is constructed for a fermion with positive energy and an arbitrary distribution of momenta. Special attention is paid to physical assumptions that should be made to enable the construction.
\end{abstract}
\section{Introduction}
In the conventional approach, the Minkowski space-time is postulated as a primary entity. Only after it, fields, differential equations of motion, etc. are defined on the space-time. One of the main aims of the relational approach to space-time geometry and physical interactions is to investigate whether it is possible to derive fundamental physical equations, most of which have been postulated based on an a priori given space-time, from some primary non-spatiotemporal relations of purely abstract mathematical character. If the answer is \textit{yes}, the next step is to find out how far one can move ahead on this way using as few physical considerations as possible, and which of conventional prerequisites are essential and which can be omitted.

Due to uncontrolled creation of particle-antiparticle pairs at $2mc^2$ energies, it is impossible to determine coordinates of a relativistic particle by collisions in principle. The particle loses its individuality. Therefore, coordinates become unobservable at the microscopic level. However, a momentum of a free particle is determined always. A remarkable discussion of that can be found in~\cite{Chew:1963}.

From the relational viewpoint, space-time is interpreted as a secondary entity emergent from more fundamental considerations, in particular, from the interactions of elementary particles. In few words, particles exist out of space-time. An interaction of two particles consists in redistribution of their quantum numbers. We mean quantum numbers expressing the physical properties of particles, i.e., coordinates and time are excluded as expressing the geometrical ones. Direct interactions of a great many of particles result in forming the classical space-time at the macroscopic level on an average. Close ideas known as the statistical interpretation of space-time can be found in the works of E.~Mach, A.~Einstein, A.~Eddington~\cite{Eddington:1946}, L.~de Broglie, J.~Wheeler, E.~Wigner~\cite{Wigner:1956}, D.~van Dantzig~\cite{Dantzig:1956}, E.~Zimmerman~\cite{Zimmerman:1962}, G.~Chew~\cite{Chew:1963}, R.~Penrose~\cite{Penrose:1977}, L.~Smolin~\cite{Smolin:2018}, C.~Rovelli, D.~Oriti~\cite{Oriti:2017} and other authors.

The most important quantum numbers are energy, momentum, and spin. We believe that all of them have the non-spatiotemporal nature. Indeed, quantum numbers are eigenvalues of Hermitian operators acting on an abstract Hilbert space. Therefore, the energy and the momentum of a quantum particle are abstract quantities too. These quantities become the energy and the momentum of classical mechanics only \textit{after} an interaction of the particle with a classical macroscopic instrument. Spin is abstract primordially and classifies finite-dimensional irreducible representations of the group $\mathrm{SU}(2)$. Thus, we choose the momentum representation of wave functions as preferable for the description of quantum particles.

From both relativistic mechanics and Feynman diagrams, we know that any interaction manifests itself in changing 4-momenta of involved particles. However, interacting particles are \textit{free} before and after an act of interaction. Thus, the first problem is to describe free particles.

In this paper, we analyze the Dirac equation for a free relativistic spin-$\frac{1}{2}$ particle from the relational viewpoint of \textit{binary pregeometry}~\cite{Vladimirov_kniga:2020}. Binary pregeometry uses a mathematical formalism of \textit{binary systems of complex relations} (BSCRs). BSCRs provide an interesting method to define coordinates in abstract sets. This method is purely algebraic and can be applied to gravitation~\cite{Vladimirov:2008}, astrophysics~\cite{Vlad_Bol_Bab:2018}, and cosmology~\cite{Molchanov:2022}. Our approach is ideologically similar to the ``twistor programme''~\cite{Penrose:1977} by R.~Penrose in which the points of the Minkowski space-time are represented by 2-dimensional linear subspaces of a complex 4-dimensional vector space with a Hermitian form of signature $(++--)$. Twistors~\cite{Penrose:1967} lead to the conformal group of the 4-dimensional pseudo-Euclidean space while 2-spinors lead to the Lorentz group. The simplest BSCR allows us to obtain from it the 2-spinor space, symplectic and unitary scalar multiplications of 2-spinors simultaneously. The main idea of the paper is to combine those scalar multiplications into a Hermitian generalization of the known Hodge operator from the exterior algebra. This generalization generates a Dirac wave packet with positive energy and arbitrary distribution of momenta of a free fermion. The Dirac equation emerges as an identity for that wave packet. We construct the wave function (wave packet) of a free particle, not the Dirac equation as such. We do not modify the Dirac equation. We try simply to look at it from the alternative (relational) viewpoint.

In Section 2, we recall a definition of rank-$(3,3)$ BSCR and deduce the 2-spinor algebra from it. This BSCR brings not only the symplectic and unitary scalar multiplications of 2-spinors, but also the groups $\mathrm{SL}(2,\mathbf{C})$ and $\mathrm{U}(2)$, which are very important for relativistic quantum mechanics. In particular, we construct a vector of a 4-dimensional pseudo-Euclidean space from 2-spinor components. This vector is proportional to a 4-momentum of a free spin-$\frac{1}{2}$ particle. BSCRs of higher ranks lead to pseudo-Finslerian generalizations of spinors~\cite{Solo_Vlad:2001,Solov'yov:2015}.

In Section 3, we discuss the prerequisites leading to the Dirac equation for a free massive spin-$\frac{1}{2}$ particle and construct its wave function in a quantum state with positive energy and an arbitrary distribution of momenta. We show that the Dirac equation emerges as an $\mathrm{SL}(2,\mathbf{C})$ and P invariant relation, connecting bispinor components of the wave function and based on the Hermitian generalization of the Hodge operator.

In Conclusion, we summarize the results of the study and discuss further generalizations of the developed formalism, in particular, for the case of interacting spin-$\frac{1}{2}$ fermions.
\section{Rank-(3, 3) BSCR and 2-spinor algebra}
Let us begin with the definition of rank-$(3,3)$ BSCR~\cite{Vladimirov_kniga:2020}. This BSCR includes (a) two sets $\mathcal{M}=\{i,k,j,\dots\}$ and $\mathcal{N}=\{\alpha,\beta,\gamma,\dots\}$ of elements of arbitrary nature; (b) a complex-valued mapping $u\colon\mathcal{M}\times\mathcal{N} \to\mathbf{C},\; (i,\alpha)\mapsto u_{i\alpha}$; (c) the requirement
\begin{equation}
\left|
\begin{array}{ccc}
u_{i\alpha}&u_{i\beta}&u_{i\gamma}\\
u_{k\alpha}&u_{k\beta}&u_{k\gamma}\\
u_{j\alpha}&u_{j\beta}&u_{j\gamma}
\end{array}
\right|=0
\label{eq:1}
\end{equation}
for \textit{any} elements $i,k,j\in\mathcal{M}$ and $\alpha,\beta,\gamma\in\mathcal{N}$. Such a construction allows us to characterize elements of $\mathcal{M}$ and $\mathcal{N}$ by numerical parameters. Indeed, let us fix the elements $k,j\in\mathcal{M}$ and $\beta,\gamma\in\mathcal{N}$ so that $u_{k\beta}=u_{j\gamma}=1$ and $u_{j\beta}=u_{k\gamma}=0$. Using Eq.~(\ref{eq:1}) and the notation $i^1=u_{i\beta}$,  $i^2=u_{i\gamma}$, $\alpha^1=u_{k\alpha}$, $\alpha^2=u_{j\alpha}$, we find
\begin{equation}
u_{i\alpha}=i^1\alpha^1+i^2\alpha^2.
\label{eq:2} 
\end{equation}
Thus, $u_{i\alpha}$ is a homogeneous polynomial of degree 2 in the complex parameters $i^1$, $i^2$, $\alpha^1$, $\alpha^2$ for any elements $i\in\mathcal{M}$ and $\alpha\in\mathcal{N}$.

In general, there is no a one-to-one correspondence between, for example, elements $i\in\mathcal{M}$ and ordered pairs $(i^1,i^2)\in\mathbf{C}^2$ of their parameters. To avoid this situation, we should specify the mapping $u\colon\mathcal{M}\times\mathcal{N}\to\mathbf{C}$. It is sufficient to assume that $\mathcal{M}$ and $\mathcal{N}$ are 2-dimensional vector spaces over the field $\mathbf{C}$ of complex numbers. The last assumption is most suitable for aims of this paper. Hereinafter, parameters of elements are components of vectors with respect to some bases in $\mathcal{M}$ and $\mathcal{N}$.

It is readily seen that Eq.~(\ref{eq:1}) is invariant under arbitrary linear transformations
\begin{equation}
i^{\prime r}=C^r_si^s\quad\hbox{and}\quad\alpha^{\prime r}=\widetilde C^r_s\alpha^s,
\label{eq:3}
\end{equation}
where $C^r_s$, $\widetilde C^r_s\in\mathbf{C}$ ($r,s=1,2$) and $s$ is a summation index as usual. Indeed, recalling the basic properties of the determinant and using Eq.~(\ref{eq:2}) for all elements, one can verify the validity of Eq.~(\ref{eq:1}) both before and after transformations~(\ref{eq:3}).

In binary pregeometry, an important role belongs to $2\times 2$ minors of determinant~(\ref{eq:1}). Those are referred to as fundamental $2\times 2$ relations and predetermine the pseudo-Euclidean character of the resulting geometry.

Let us consider the fundamental $2\times 2$ relation between elements $(i,k)\in\mathcal{M}^2$ and $(\alpha,\beta)\in\mathcal{N}^2$ and, first of all, rewrite it in the form
\begin{equation}
\left[{\alpha\atop i}{\beta\atop k}\right]\equiv
\left|\begin{array}{cc}
u_{i\alpha}&u_{i\beta}\\
u_{k\alpha}&u_{k\beta}
\end{array}\right|=
\left|\begin{array}{cc}
i^1&k^1\\
i^2&k^2
\end{array}\right|
\left|\begin{array}{cc}
\alpha^1&\beta^1\\
\alpha^2&\beta^2
\end{array}\right|.
\label{eq:4}
\end{equation}
Now, we select such transformations from Eq.~(\ref{eq:3}) that leave fundamental $2\times 2$ relation~(\ref{eq:4}) unchanged. Since coefficients $C^r_s$ and $\widetilde C^r_s$ are mutually independent, it suffices only to require the invariance of each of the two determinants
\begin{equation}
[i,k]\equiv
\left|\begin{array}{cc}
i^1&k^1\\
i^2&k^2
\end{array}\right|=\varepsilon_{rs}i^rk^s
\quad\hbox{and}\quad
[\alpha,\beta]\equiv
\left|\begin{array}{cc}
\alpha^1&\beta^1\\
\alpha^2&\beta^2
\end{array}\right|=\varepsilon_{rs}\alpha^r\beta^s.
\label{eq:5}
\end{equation}
Here, $\varepsilon_{rs}$ is the 2-dimensional Levi-Civita symbol. The last requirement can be readily satisfied by using the following formulas
\begin{equation}
\left|\begin{array}{cc}
i^{\prime1}&k^{\prime1}\\
i^{\prime2}&k^{\prime2}
\end{array}\right|=
\det C_2
\left|\begin{array}{cc}
i^1&k^1\\
i^2&k^2
\end{array}\right|
\quad\hbox{and}\quad
\left|\begin{array}{cc}
\alpha^{\prime1}&\beta^{\prime1}\\
\alpha^{\prime2}&\beta^{\prime2}
\end{array}\right|=
\det\widetilde C_2
\left|\begin{array}{cc}
\alpha^1&\beta^1\\
\alpha^2&\beta^2
\end{array}\right|,
\label{eq:6}
\end{equation}
where $C_2=\|C^r_s\|$ and $\widetilde C_2=\|\widetilde C^r_s\|$ are $2\times 2$ complex matrices. Indeed, it follows straightforward from Eq.~(\ref{eq:6}) that determinants~(\ref{eq:5}) are invariant if and only if the matrices $C_2$ and $\widetilde C_2$ are unimodular, i.e.,
\begin{equation}
\det C_2=\det\widetilde C_2=1.
\label{eq:7} 
\end{equation}
Thus, the desired transformations form the group isomorphic to $\mathrm{SL}(2,\mathbf{C})$ due to conditions~(\ref{eq:7}).

The expressions $[i,k]=\varepsilon_{rs}i^rk^s$ and $[\alpha,\beta]=\varepsilon_{rs} \alpha^r\beta^s$ in Eq.~(\ref{eq:5}) are antisymmetric bilinear forms on $\mathbf{C}^2$, and those are nondegenerate, since $\det\|\varepsilon_{rs}\|=1\not=0$. Up to an isometry, those are the unique symplectic scalar products in two dimensions. Hence, $(\mathcal{M};[\cdot,\cdot])$ and $(\mathcal{N};[\cdot,\cdot])$ are 2-dimensional symplectic spaces over $\mathbf{C}$ with the Levi-Civita symbol as a metric tensor, which can be used to lower indices, e.g., $i_r=\varepsilon_{rs}i^s$ and $\alpha_r=\varepsilon_{rs}\alpha^s$. The isometry group of these spaces is isomorphic to $\mathrm{SL}(2,\mathbf{C})$.

Let us recall that a 2-spinor (spin-vector)~\cite{Penrose_Rindler:1984} is defined traditionally as an element of a 2-dimensional vector space over $\mathbf{C}$ equipped with the symplectic scalar multiplication. But $(\mathcal{M};[\cdot,\cdot])$ and $(\mathcal{N};[\cdot,\cdot])$ belongs to such a type of spaces exactly. Therefore, the parameters of elements of $\mathcal{M}$ and $\mathcal{N}$ can be viewed as the components of 2-spinors.

Up to this point we assumed $\mathcal{M}$ and $\mathcal{N}$ to be independent of each other. However, a case of the main interest for physics is the one when those are antilinearly isomorphic as vector spaces (since one can obtain the 4-dimensional pseudo-Euclidean geometry and the Lorentz group only in this case). The antilinear isomorphism can be defined explicitly in the form
\begin{equation}
\alpha^r=\overline{i^r},\ \beta^r=\overline{k^r},\dots,
\label{eq:8}
\end{equation}
where overlines denote complex conjugation. It is clear that the matrices of transformations~(\ref{eq:3}) cannot be mutually independent anymore and their elements should satisfy the conditions $\widetilde C^r_s=\overline{C^r_s}$. Below, quantities under transformation with the matrix $\overline C_2=\|\overline{C_s^r}\|$ will have dotted indices (the van der Waerden notation~\cite{Waerden:1928}).

As we have noted above, $(\mathcal{M};[\cdot,\cdot])$ is the space of 2-spinors. Then, the space $(\mathcal{N};[\cdot,\cdot])$, for which according to Eq.~(\ref{eq:8}) we have $[\alpha,\beta]=\overline{[i,k]}$, is the space of complex conjugates of 2-spinors. These two spaces generate the general spin-tensor algebra~\cite{Penrose_Rindler:1984}.

Let us consider the pair relation $u_{i\beta}$ between elements $i\in\mathcal{M}$ and $\beta\in\mathcal{N}$. Replacing $\alpha$ with $\beta$ in Eq.~(\ref{eq:2}) and using Eq.~(\ref{eq:8}), we can write $u_{i\beta}$ as
\begin{equation}
u_{i\beta}=i^1\beta^{\dot 1}+i^2\beta^{\dot 2}=i^1\overline{k^{\dot 1}}+i^2\overline{k^{\dot 2}}
\equiv\langle i,k\rangle.
\label{eq:9}
\end{equation}
The algebraic form $\langle i,k\rangle=\delta_{r\dot s}i^r\overline{k^{\dot s}}$, where $\delta_{r\dot s}$ is the Kronecker symbol, is sesquilinear (linear with respect to the first argument and antilinear with respect to the second one: $\langle\lambda i,\rho
k\rangle=\lambda\overline\rho\langle i,k\rangle$; $\lambda,\rho\in\mathbf{C}$), Hermitian ($\langle k,i\rangle=\overline{\langle i,k\rangle}$), and positive definite ($\langle i,i\rangle>0$ for $i^r\not=0$). Therefore, Eq.~(\ref{eq:9}) can be interpreted as a unitary scalar product of elements $i$ and $k$ so that $(\mathcal{M};\langle\cdot,\cdot\rangle)$ is a 2-dimensional unitary space. The same is obviously true for the space $(\mathcal{N};\langle\cdot,\cdot\rangle)$, in which $\langle\alpha,\beta\rangle=\overline{\langle i,k\rangle}$ according to Eqs.~(\ref{eq:8}) and (\ref{eq:9}).

With this done, we are ready to construct vectors of a 4-dimensional pseudo-Euclidean vector space with the metric signature $({+}{-}{-}{-})$ and transformations from $\mathrm{O}^{\uparrow}_{+}(1,3)$ (the proper orthochronous Lorentz group).

Let us see how it works. Following~\cite{Vladimirov_kniga:2020}, we first consider a contravariant spin-tensor of a special form with one ordinary and one dotted indices:
\begin{equation}
V^{r\dot s}=i^r\alpha^{\dot s}+k^r\beta^{\dot s}.
\label{eq:10}
\end{equation}
Let $V_2=\|V^{r\dot s}\|$ be the matrix of spin-tensor components~(\ref{eq:10}). Then, from Eqs.~(\ref{eq:8}) and (\ref{eq:10}), we have $V^{r\dot s}=\overline{V^{s\dot r}}$ or, which is the same, $V_2=V_2^{+}$ (hereinafter, the cross denotes Hermitian conjugation). Hence, the matrix $V_2$ is Hermitian.

It is clear that $V^{1\dot1}=\left|i^1\right|^2+\left|k^1\right|^2\geq 0$. Moreover, in view of Eqs.~(\ref{eq:10}), (\ref{eq:4}), (\ref{eq:5}), and~(\ref{eq:8}), 
\begin{equation}
\det V_2=\left[{\alpha\atop i}{\beta\atop k}\right]=\Bigl|[i,k]\Bigr|^2
\geq 0.
\label{eq:11}
\end{equation} 
However, fundamental $2\times 2$ relation~(\ref{eq:11}) is strictly positive if and only if the elements $i$ and $k$ are linearly independent. Thus, if $i$ and $k$ form a basis in $\mathcal{M}$, then  $V^{1\dot1}>0$ and $\det V_2 >0$, i.e., according to the Sylvester criterion, the matrix $V_2$ is positive definite.

As we found above, $V_2=V^{+}_2$. In other words, $V_2$ is an element of the 4-dimensional vector space $\mathrm{Herm}(2)$ over the field $\mathbf{R}$ of real numbers which is formed by all $2\times 2$ Hermitian matrices. Let us choose a basis in $\mathrm{Herm}(2)$ consisting of the unit matrix $\sigma_0$ and the Pauli matrices $\sigma_1$, $\sigma_2$, $\sigma_3$:
\begin{equation}
\sigma_0=
\left(\begin{array}{cc}
1&0\\
0&1
\end{array}\right),\
\sigma_1=
\left(\begin{array}{cc}
0&1\\
1&0
\end{array}\right),\
\sigma_2=
\left(\begin{array}{cc}
0&-\iota\\
\iota&0
\end{array}\right),\
\sigma_3=
\left(\begin{array}{cc}
1&0\\
0&-1\end{array}\right).
\label{eq:12}
\end{equation}
Here $\iota$ is the imaginary unit ($\iota^2=-1$). Therefore, we can decompose $V_2\in\mathrm{Herm}(2)$ as
\begin{equation}
V_2=v^\mu\sigma_\mu\quad(\mu=0,1,2,3),
\label{eq:13}
\end{equation}
where $v^0,\dots,v^3\in\mathbf{R}$ are components of $V_2$ with respect to the basis $\{\sigma_\mu\}$. Using Eq.~(\ref{eq:13}), we represent $v^\mu$ directly in terms of spin-tensor components~(\ref{eq:10}):
\begin{eqnarray}
v^0=\frac{1}{2}(V^{1\dot1}+V^{2\dot2})=\frac{1}{2}
(i^1\alpha^{\dot1}+i^2\alpha^{\dot2}+k^1\beta^{\dot1}+k^2\beta^{\dot2}),
\nonumber\\
v^1=\frac{1}{2}(V^{1\dot2}+V^{2\dot1})=\frac{1}{2}
(i^1\alpha^{\dot2}+i^2\alpha^{\dot1}+k^1\beta^{\dot2}+k^2\beta^{\dot1}),
\nonumber\\
v^2=\frac{\iota}{2}(V^{1\dot2}-V^{2\dot1})=\frac{\iota}{2}
(i^1\alpha^{\dot2}-i^2\alpha^{\dot1}+k^1\beta^{\dot2}-k^2\beta^{\dot1}),
\nonumber\\
v^3=\frac{1}{2}(V^{1\dot1}-V^{2\dot2})=\frac{1}{2}
(i^1\alpha^{\dot1}-i^2\alpha^{\dot2}+k^1\beta^{\dot1}-k^2\beta^{\dot2}).
\label{eq:14}
\end{eqnarray}
Using the well-known formula $\mathrm{tr}(\sigma_\mu\sigma_\nu)=
2\delta_{\mu\nu}$ and the notation $\sigma^\mu\equiv\sigma_\mu$, $\mathsf{i}\equiv(i^1,i^2)^\top$, $\mathsf{k}\equiv(k^1,k^2)^\top$ ($\scriptstyle\top$ is matrix transposition), we can rewrite Eq.~(\ref{eq:14}) in the compact form:
$$
v^\mu=\frac{1}{2}\mathrm{tr}(\sigma^\mu V_2)=\frac{1}{2}(\mathsf{i}^{+}
\sigma^\mu\mathsf{i}+\mathsf{k}^{+}\sigma^\mu\mathsf{k}).\eqno(14^\prime)
$$
In what follows, we will find conditions under which the vector $v^\mu$ with components~(\ref{eq:14}) coincides with the 4-momentum of a free spin-$\frac{1}{2}$ fermion.

Let us calculate the determinant of the left and right sides of Eq.~(\ref{eq:13}). It is evident that $\det V_2=(v^0)^2-(v^1)^2-(v^2)^2-(v^3)^2$. On the other hand, we have formula~(\ref{eq:11}). Thus, fundamental $2\times 2$ relation~(\ref{eq:4}) is nothing else but a pseudo-Euclidean scalar square of the 4-vector $v^\mu$:
\begin{equation}
\left[{\alpha\atop i}{\beta\atop k}\right]=g_{\mu\nu}v^\mu v^\nu\equiv v^2
\geq 0,
\label{eq:15}
\end{equation}
where $\|g_{\mu\nu}\|=\mathrm{diag}(1,-1,-1,-1)$ is a metric tensor of the space $\mathbf{R}^4_{1,3}$. Moreover, it follows from Eq.~(\ref{eq:15}) that $v^\mu$ is either timelike (if $i$ and $k$ are linearly independent) or isotropic (if $i$ and $k$ are linearly dependent). Since, in addition, $v^0=\frac{1}{2}(
\mathsf{i}^{+}\mathsf{i}+\mathsf{k}^{+}\mathsf{k})>0$, the vector $v^\mu$ belongs to the upper part of the isotropic cone in $\mathbf{R}^4_{1,3}$.

Linear transformations~(\ref{eq:3}) with conditions~(\ref{eq:8}) induce the transformation $V^{\prime r\dot s}=C^r_t\overline{C^{\dot s}
_{\dot u}}V^{t\dot u}$ of spin-tensor~(\ref{eq:10}), where $V^{\prime r\dot s}=i^{\prime r}\alpha^{\prime \dot s}+k^{\prime r}\beta^{\prime \dot s}$. It should be noted that $V^{t\dot u}$ and $V^{\prime r\dot s}$ are \textit{different} spin-tensors, i.e., the above transformation is an active one. It can be represented in the matrix form
\begin{equation}
V_2^\prime=C_2 V_2 C_2^{+}\equiv\widehat{L}(C_2)V_2\quad(V^\prime_2=\|V^{\prime r\dot s}\|),
\label{eq:16}
\end{equation}
where the operator $\widehat{L}(C_2)\colon V_2\mapsto V_2^\prime$ acts on $2\times 2$ complex matrices and has the following obvious properties:
\begin{enumerate}
\item $\widehat{L}(C_2)$  is linear.
\item $\widehat{L}(C_2)$ transforms Hermitian matrices into Hermitian ones.
\item $\widehat{L}(C_2)\widehat{L}(\widetilde{C}_2)=\widehat{L}(C_2\widetilde{C}
_2)$ for any $2\times 2$ complex matrices $C_2$ and $\widetilde{C}_2.$
\end{enumerate}
From (a) and (b) we conclude that Eq.~(\ref{eq:16}) is a linear transformation of the space $\mathrm{Herm}(2)$. 

Using Eq.~(\ref{eq:12}), we can rewrite Eq.~(\ref{eq:16}) in the form
\begin{equation}
v^{\prime\mu}=L(C_2)^{\mu}_\nu v^\nu\quad(\mu,\nu=0,1,2,3),
\label{eq:17}
\end{equation}
where $L(C_2)=\|L(C_2)^\mu_\nu\|$ is the matrix of the operator $\widehat{L}(C_2)$ with respect to the basis $\{\sigma_\mu\}$. Replacing $V_2$ and $V^\prime_2$ in Eq.~(\ref{eq:16}) by their decompositions into basis~(\ref{eq:12}) and using the formula $\mathrm{tr}(\sigma^\mu\sigma_\nu)=2\delta^\mu_\nu$, we get the real-valued elements 
\begin{equation}
L(C_2)^\mu_\nu=\frac{1}{2}\mathrm{tr}(\sigma^\mu C_2\sigma_\nu
C^{+}_2)
\label{eq:18}
\end{equation}
of the above $4\times 4$ matrix $L(C_2)$. It is seen from Eq.~(\ref{eq:18}) that $L(C_2)_0^0>0$ for any nonzero $2\times 2$ complex matrix $C_2$.

Computing the determinant of the left and right sides of Eq.~(\ref{eq:16}) gives $\det
V_2^\prime=|\det C_2|^2\det V_2$. However, as it was shown above, $\det
V_2=g_{\mu\nu}v^\mu v^\nu$ (similarly, $\det V_2^\prime=g_{\mu\nu}v^{\prime\mu}
v^{\prime\nu}$). Hence,
\begin{equation}
g_{\mu\nu}v^{\prime\mu}v^{\prime\nu}=|\det C_2|^2 g_{\mu\nu}v^\mu v^\nu,
\label{eq:19}
\end{equation}
i.e., for $C_2\in\mathrm{GL}(2,\mathbf{C})$ transformation~(\ref{eq:17}) is a conformal one.

Let us assume $\det C_2=1$. Then, according to Eq.~(\ref{eq:19}) $g_{\mu\nu}v^{\prime\mu}v^{\prime\nu}=g_{\mu\nu}v^\mu v^\nu$. This means that, for any $C_2\in\mathrm{SL}(2,\mathbf{C})$, transformation~(\ref{eq:17}) belongs to the Lorentz group $\mathrm{O}(1,3)$. Moreover, the mapping $C_2\mapsto L(C_2)$ which sends a $2\times 2$ complex matrix with unit determinant to a $4\times 4$ real matrix of the linear operator $\widehat L(C_2)$ in the basis $\{\sigma_\mu\}$ is the known epimorphism (surjective homomorphism) from the group $\mathrm{SL}(2,\mathbf{C})$ onto the group $\mathrm{O}^\uparrow_+(1,3)$~\cite{Penrose_Rindler:1984}. This mapping is homomorphic due to the property (c) of the operator $\widehat L(C_2)$.

\section{Dirac equation for free fermion in momentum representation}
According to the previous section, the vector space $\mathcal M$ is equipped with the symplectic scalar multiplication $[\cdot,\cdot]$ and the unitary scalar multiplication $\langle\cdot,\cdot\rangle$ simultaneously. These scalar multiplications will allow us to define a set of antilinear automorphisms which generates the Dirac equation for a free massive spin-$\frac{1}{2}$ fermion in the momentum representation. However, let us begin with some \textit{physical} assumptions.

First of all, note one circumstance of general character. Every particle possesses physical properties (energy, momentum, spin, charge) and geometrical ones (space-time position). It is natural to base quantum description of a particle on its physical properties, because those are measured in experiments. For this reason, we will characterize a state of a free massive spin-$\frac{1}{2}$ particle by a wave function in the \textit{momentum representation}.
  
We have already introduced 2-spinors and the unitary scalar multiplication $\langle\cdot,\cdot\rangle$. Let us interpret spinor components $i^r_\circ$ ($r=1,2$) as probability amplitudes to find a fermion in one of the two possible spin states. Of course, the sum of the probabilities of all the possible spin states is equal to one: $\langle i_\circ,i_\circ\rangle=|i_\circ^1|^2+|i_\circ^2|^2=1$ (the fermion must be found in one of the two spin states).

However, we still know nothing about the probability amplitude for the fermion to have a given momentum. The appropriate wave function should be complex-valued, defined on the 3-dimensional momentum space, and have two components corresponding to the two spin states. First of all, we need to construct the momentum space of a free massive fermion. This can be readily done by using the results of the previous section.

We saw above that $[i,k]=\varepsilon_{rs}i^rk^s$ is invariant under $\mathrm{SL}(2,\mathbf{C})$ transformations~(\ref{eq:3}) which induce Lorentz transformations of the space $\mathrm{Herm}(2)$. It is evident that $\langle i_\circ,i_\circ\rangle=\delta_{r\dot s}i_\circ^r\overline{i_\circ^{\dot s}}$ is \textit{not invariant} under such transformations (with the exception of $\mathrm{SU}(2)$ transformations). In order to save the quantum probabilistic interpretation, we suppose that the unitary geometry on $\mathcal{M}$ is defined up to an isomorphism. In other words, we identify an $\mathrm{SL}(2,\mathbf{C})$ transformation $i^r=C^r_s i_\circ^s$ with an isomorphism between the unitary spaces $(\mathcal{M};\langle\cdot,\cdot\rangle)$ and $(\mathcal{M};\langle\cdot,\cdot\rangle_u)$, where $\langle\cdot,\cdot\rangle_u$ is another unitary scalar multiplication which satisfies the requirement $\langle i,i\rangle_u=\langle i_\circ,i_\circ\rangle$ for any $i_\circ\in\mathcal{M}$.

If $\langle i,i\rangle_u=U_{r\dot s}i^r\overline{i^{\dot s}}$, then this requirement implies $U_{r\dot s}=(C_2^{-1})^p_r\overline{(C_2^{-1})^{\dot q}_{\dot s}}\delta_{p\dot q}$ or, in the matrix form, $U_2=(C_2^{-1})^\top\overline{C_2^{-1}}$ ($U_2=\|U_{r\dot s}\|$, $C_2=\|C^r_s\|$). Therefore, $U_2\in\mathrm{Herm}(2)$ and is a positive definite matrix, to which the constructions of the previous section can be applied (with replacement of superscripts by subscripts and vice versa). In particular, we can write
\begin{equation}
U_2=u_\mu \sigma^\mu,
\label{eq:20} 
\end{equation}
where $u_\mu$ is a 4-dimensional time-like covector with $u_0>0$. Since $C_2\in\mathrm{SL}(2,\mathbf{C})$, we have $\det U_2=\det[ (C_2^{-1})^\top\overline{C_2^{-1}}]=1$. On the other hand, because of Eq.~(\ref{eq:20}), $\det U_2=g^{\mu\nu}u_\mu u_\nu$, where $\|g^{\mu\nu}\|=\mathrm{diag}(1,-1,-1,-1)$ is the contravariant metric tensor of the space $\mathbf{R}^4_{1,3}$. Therefore, $g^{\mu\nu}u_\mu u_\nu=1$. This result substantiates the interpretation of $u_\mu$ as a 4-velocity of a massive fermion. It is not difficult to show that $\mathbf{u}=(u_1,u_2,u_3)$ runs $\mathbf{R}^3$ when $C_2$ runs the subset of Hermitian positive definite matrices in the group $\mathrm{SL}(2,\mathbf{C})$ (Lorentz boosts). Thus, we have expressed the 4-velocity $u_\mu$ of the fermion in terms of the parameters $C^r_s$ of the $\mathrm{SL}(2,\mathbf{C})$ transformation. The same is valid for a 4-momentum $p_\mu$ of a fermion with a mass $m$ because $p_\mu=mu_\mu$ by definition (in the system of units, where the speed of light $c=1$). Notice, that $p_0=\sqrt{\mathbf{p}^2+m^2}>0$ automatically for any $\mathbf{p}\in\mathbf{R}^3$ ($u_0>0$ and $g^{\mu\nu}u_\mu u_\nu=1$).

Now, we can define a natural antilinear automorphism $i\in\mathcal{M}\mapsto k\in\mathcal{M}$ of the space $(\mathcal{M};[\cdot,\cdot], \langle\cdot,\cdot\rangle_u)$ by the following formula
\begin{equation}
i^r\mapsto\overline{k^{\dot u}}=\varepsilon^{\dot s\dot u}U_{r\dot s}i^r,
\label{eq:21} 
\end{equation}
where $\varepsilon^{\dot s\dot u}$ are contravariant components of the metric spin-tensor $\varepsilon_{\dot r\dot u}$ such that $\varepsilon_{\dot r\dot u}\varepsilon^{\dot s\dot u}=\delta_{\dot r}^{\dot s}$. Automorphism~(\ref{eq:21}) is fully analogous to the known Hodge operator~\cite{Flanders:1963}. Because of Eq.~(\ref{eq:20}) and the equality $u_0=\sqrt{\mathbf{u}^2+1}$, $U_{r\dot s}$ are the functions of $\mathbf{u}\in\mathbf{R}^3$, i.e., we actually have a 3-parametric family of antilinear automorphisms~(\ref{eq:21}). For each $\mathbf{u}\in\mathbf{R}^3$, we choose one element $i(\mathbf{u})\in\mathcal{M}$. An arbitrary 2-spinor field $i^r(\mathbf{u})$ appears in this way. Inserting $i^r=i^r(\mathbf{u})$ into Eq.~(\ref{eq:21}) and using Eq.~(\ref{eq:20}), we obtain the remarkable relation
\begin{equation}
\beta_{\dot s}=(u_\mu\sigma^\mu)_{r\dot s}i^r(\mathbf{u}),
\label{eq:22} 
\end{equation}
where $\beta_{\dot r}=\varepsilon_{\dot r\dot u}\overline{k^{\dot u}}$. In fact, Eq.~(\ref{eq:22}) is equivalent to the Dirac equation for a free massive spin-$\frac{1}{2}$ fermion with positive energy in the momentum representation. Our final step is to show that it is really so.

We proceed with this after making another important assumption. It concerns the property of mirror symmetry, or P invariance. It is known that processes observed in inanimate nature (with the exception of those related to weak interactions) are left-right symmetric. In particular, the mirror reflection of an allowed wave function should transform it into another allowed one. The P reflection (space inversion) acts on an arbitrary true vector  $v^\mu$ according to the rule: $v^0$, $v^1$, $v^2$,
$v^3\mapsto v^0$, $-v^1$, $-v^2$, $-v^3$. Then, what is the rule in the case of 2-spinors? 
       
Clearly, there are no P reflections in the group $\mathrm{O}^\uparrow_+(1,3)$ and, therefore, $\mathrm{SL}(2,\mathbf{C})$ transformations. However, formulas~(\ref{eq:14}) show that the P reflection of the vector $v^\mu$ is equivalent to the following transformation of spin-tensor components: $V^{1\dot 1}\mapsto V^{2\dot 2}$,
$V^{2\dot 2}\mapsto V^{1\dot 1}$, $V^{1\dot 2}\mapsto -V^{1\dot 2}$,
$V^{2\dot 1}\mapsto -V^{2\dot 1}$. This can be represented in a more compact form: $V^{r\dot s}\mapsto V_{s\dot r}$, where $V_{s\dot r}=\varepsilon_{st}\varepsilon_{\dot r\dot u}V^{t\dot u}$. Twice repeated, the P reflection should return the vector $v^\mu$ or the equivalent spin-tensor $V^{r\dot s}$  into the initial (non-transformed) state, i.e., $V_{s\dot r}\mapsto V^{r\dot s}$. Therefore, if we want to have objects that are transformed into themselves under the P reflection, we should introduce ordered pairs of spin-tensors with different valences $(V^{r\dot s},V_{s\dot r})$. Such pairs are transformed according to the rule: $(V^{r\dot s},V_{s\dot r})\mapsto(V_{s\dot r},V^{r\dot s})$. In other words, the P reflection swaps elements in the pair.

It is natural to expect similar behavior of spinors under the P reflection. Let us specify the P reflection in the form: $(i^r,\beta_{\dot r})\mapsto(\beta_{\dot r},i^r)$, $(k^r,\alpha_{\dot r})\mapsto(\alpha_{\dot r},k^r)$, where $\alpha_{\dot r}=\varepsilon_{\dot r\dot u}\alpha^{\dot u}$, $\beta_{\dot r}=\varepsilon_{\dot r\dot u}\beta^{\dot  u}$. The validity of this choice stems from Eq.~(\ref{eq:8}) and formulas (\ref{eq:14}) for the components of the vector $v^\mu$ which behave precisely as required  under the P reflection: $v^0$, $v^1$, $v^2$, $v^3\mapsto v^0$, $-v^1$, $-v^2$, $-v^3$. As far as, in the relational approach, vectors are secondary concepts constructed from 2-spinors, it is more appropriate to say that the P reflection of vectors arises as a representation of the P reflection of spinors.  

The mapping $(i^r,\beta_{\dot r})\mapsto(\beta_{\dot r},i^r)$ is not the only way to define the P reflection of spinors, although it is the most simple one. Alternative variants result from multiplying it by a complex factor of unit modulus~\cite{Berestetskii:1982}.

Let us consider the relation which is \textit{inverse} with respect to Eq.~(\ref{eq:22}). It has the form
\begin{equation}
i^r(\mathbf{u})=(g^{\mu\nu}u_\mu\overline{\sigma}_\nu)^{r\dot s}\beta_{\dot s},
\label{eq:23} 
\end{equation}
where $\overline{\sigma}_\mu\equiv\overline{\sigma_\mu}$ and $U^{r\dot s}=(g^{\mu\nu}u_\mu\overline{\sigma}_\nu)^{r\dot s}$ are contravariant components of the unitary metric tensor $U_{r\dot s}=(u_\mu\sigma^\mu)_{r\dot s}$ such that $U_{r\dot s}U^{u\dot s}=\delta^u_r$. Notice, that Eq.~(\ref{eq:23}) is generated by the inverse antilinear automorphism $k^u\mapsto i^r=U^{r\dot s}\varepsilon_{\dot s\dot u}\overline{k^{\dot u}}$ of the space $(\mathcal{M};[\cdot,\cdot], \langle\cdot,\cdot\rangle_u)$. Applying the P reflection ($i^r\mapsto\beta_{\dot r}$, $\beta_{\dot s}\mapsto i^s$, $U^{r\dot s}\mapsto U_{s\dot r}$, $U_{s\dot r}\mapsto U^{r\dot s}$) to all members of Eqs.~(\ref{eq:22})--(\ref{eq:23}), we see that the above equations pass into each other. Thus, a system of the equations
\begin{eqnarray}
(g^{\mu\nu}u_\mu\overline{\sigma}_\nu)^{r\dot s}\beta_{\dot s}=i^r(\mathbf{u}),
\label{eq:24}\\
(u_\mu\sigma^\mu)_{r\dot s}i^r(\mathbf{u})=\beta_{\dot s}
\label{eq:25}
\end{eqnarray}
is \textit{invariant} under the P reflection. 

Let us recall the definition of the 4-momentum $p_\mu=mu_\mu$ and introduce the following matrices
\begin{equation}
\gamma^0=\left(\begin{array}{cc}
0&\sigma^0\\
\sigma^0&0
\end{array}\right),
\quad
\gamma^k=\left(\begin{array}{cc}
0&-\overline{\sigma}^k\\
\overline{\sigma}^k&0
\end{array}\right),
\quad
\psi(\mathbf{p})=\left(\begin{array}{c}
i^1(\mathbf{p})\\
i^2(\mathbf{p})\\
\beta_{\dot 1}\\
\beta_{\dot 2}
\end{array}\right),
\label{eq:26}
\end{equation}
where $k=1,2,3$ and $\overline{\sigma}^\mu\equiv\overline{\sigma}_\mu$. In notation~(\ref{eq:26}), the system of equations~(\ref{eq:24})--(\ref{eq:25}) is rewritten as
\begin{equation}
(p_\mu\gamma^\mu-m)\psi(\mathbf{p})=0
\label{eq:27}
\end{equation}
(we have used the evident equalities $\sigma_\mu^\top=\overline{\sigma_\mu}$). Of course, $\gamma^\mu\gamma^\nu+\gamma^\nu\gamma^\mu=2g^{\mu\nu}$ and $p_0=\sqrt{\mathbf{p}^2+m^2}$. Eq.~(\ref{eq:27}) is the traditional form of the Dirac equation for a free massive spin-$\frac{1}{2}$ fermion with positive energy in the momentum representation.

It should be noted that Eqs.~(\ref{eq:25})--(\ref{eq:26}) give the explicit expression
\begin{equation}
\psi(\mathbf{p})=
\left(\begin{array}{r}
\hat{i}(\mathbf{p})\\
\displaystyle\frac{p_\mu\overline{\sigma}^\mu}{m}\hat{i}(\mathbf{p})
\end{array}\right),\quad
\hat{i}(\mathbf{p})=
\left(\begin{array}{c}
i^1(\mathbf{p})\\
i^2(\mathbf{p})\\
\end{array}\right),\quad
\mathbf{p}\in\mathbf{R}^3
\label{eq:28}
\end{equation}
for the Dirac wave function $\psi(\mathbf{p})$, where $i^r(\mathbf{p})$ is an arbitrary 2-spinor field. Wave functions of type~(\ref{eq:28}) are transformed according to one of the irreducible unitary representations of the Poincar\'{e} group~\cite{Wigner:1948}.

Thus, beginning with antilinear automorphism~(\ref{eq:21}), we have obtained not only wave function~(\ref{eq:28}), but also Dirac equation~(\ref{eq:27}) in the momentum representation. It is not difficult to show with the help of Eq.~(\ref{eq:27}) that $p^\mu=v^\mu$ if $\psi(\mathbf{p})^{+}\gamma^0\psi(\mathbf{p})=2m$ and $v^\mu$ is vector~(\ref{eq:14}).
\section{Conclusion}
A number of final remarks should be made to summarize the results of the paper.

The role of binary systems of complex relations is following. Instead of postulating the 4-dimensional pseudo-Euclidean space, the Lorentz group, the Clifford algebra of gamma matrices, and the Dirac equation, we only postulate rank-$(3,3)$ BSCR and Hermitian generalization~(\ref{eq:21}) of the Hodge star operator. In particular, requirement~(\ref{eq:1}) admits arbitrary linear transformations~(\ref{eq:3}) of 2-spinor components. Determinant~(\ref{eq:1}) allows only two kinds of nonzero minors: $1\times 1$ minor~(\ref{eq:2}) and $2\times 2$ minor~(\ref{eq:4}). The former is interpreted as unitary scalar product~(\ref{eq:9}), while the latter generates symplectic scalar products~(\ref{eq:5}) in the 2-spinor space. The natural question on symmetry groups of the above scalar products leads us to the groups $\mathrm{U}(2)$ and $\mathrm{SL}(2,\mathbf{C})$, which are very important for relativistic quantum mechanics. $\mathrm{SL}(2,\mathbf{C})$ is locally isomorphic to the proper orthochronous Lorentz group $\mathrm{O}^\uparrow_+(1,3)$ so that the last is a consequence of rank-$(3,3)$ BSCR as well.

Our analysis confirms the opinion by S.~Weinberg: ``The free-particle Dirac equation is nothing but a Lorentz-invariant record of the convention that we have used in putting together the two irreducible representations of the proper orthochronous Lorentz group to form a field that transforms simply also under space inversion''~\cite{Weinberg:2000}.

However, the true foundation of the Dirac equation is the family of antilinear automorphisms~(\ref{eq:21}). Those lead to relation~(\ref{eq:22}) which actually defines wave functions~(\ref{eq:28}). Dirac equation~(\ref{eq:27}) emerges as an identity for these wave functions with respect to $\mathbf{p}\in\mathbf{R}^3$. The 3-dimensional momentum space of a massive fermion is generated by the 3-parameter family of Hermitian matrices in the group $\mathrm{SL}(2,\mathbf{C})$.

It is important to note that \textit{no a priori given space-time was employed}. We operated only within the 2-spinor algebra framework and managed to construct a self-consistent quantum description of a free massive fermion with positive energy. Quantum states with $p_0<0$ can be described as well. To this purpose, we should substitute $U_{r\dot s}$ with $-U_{r\dot s}$ in Eq.~(\ref{eq:21}), i.e., pass from the unitary geometry to the antiunitary one. Nevertheless, transitions $p_0\mapsto -p_0$ are forbidden for free particles so that the case $p_0>0$ is fully sufficient.

The formalism developed in this paper can be successfully applied to Finslerian $N$-spinors. In particular, for $N=3$ we obtain a 9-dimensional pseudo-Finslerian generalization of the Dirac equation~\cite{Solov'yov:2011}. It is interesting that the corresponding ``gamma matrices'' generate an algebra similar to the Duffin--Kemmer algebra.

Our approach is valid only for free fermions. For an interacting fermion, the Dirac equation in the momentum representation is integral while Eq.~(\ref{eq:27}) is purely algebraic.

One can construct ``in'' and ``out'' states of several particles from one-particle wave functions of free particles in a usual way. Later on, we can use the conventional quantum field theory for description of interactions. Namely, elements of S-matrix are calculated according to Feynman rules in scattering problems. In case of bound states, we can use integral wave equations such as the Dirac equation in the external potential for one paricle and the Bethe--Salpeter equation for two interacting particles. Those are integral equations as the result of Fourier transform from the coordinate representation to the momentum one.
\section*{Acknowledgments}
The author would like to thank Prof. Yu.S.~Vladimirov and Dr. S.V.~Bolo\-khov for numerous enlightening discussions and collaboration. The author is grateful to A.A.~Sidorova-Biryukova for the help in preparing the first English version of the paper.

\end{document}